\documentclass[aps, prd, superscriptaddress, nofootinbib, preprintnumbers, twocolumn, showpacs]{revtex4}
\usepackage{graphicx}
\usepackage{amsmath}
\def\fsl#1{\setbox0=\hbox{$#1$}                 
   \dimen0=\wd0                                 
   \setbox1=\hbox{/} \dimen1=\wd1               
   \ifdim\dimen0>\dimen1                        
      \rlap{\hbox to \dimen0{\hfil/\hfil}}      
      #1                                        
   \else                                        
      \rlap{\hbox to \dimen1{\hfil$#1$\hfil}}   
      /                                         
   \fi}                                         %

\newcommand{\NDA}{\Omega_{\rm NDA}}

\newcommand{\Ly}{\Lambda_{LY}}

\newcommand{\DLR}{\stackrel{\leftrightarrow}\partial}

\makeatletter
\@addtoreset{equation}{section}
\makeatother

\begin{document}
\title{Topcolor breaking through boundary conditions}
\date{\today}

\preprint{PNUTP-04-A09}

\preprint{MIT-CTP-3540}

\preprint{hep-ph/0409223}

\author{Michio Hashimoto}
\email[E-mail: ]{michioh@charm.phys.pusan.ac.kr, mhashimo@uwo.ca}
\email[The present address is
{\it Department of Applied Mathematics,
Western Science Centre, The University of Western Ontario,
London ON Canada N6A 5B7}]{}
\affiliation{Department of Physics, Pusan National University,
             Pusan 609-735, Korea}
\author{Deog Ki Hong}
\email[E-mail: ]{dkhong@pusan.ac.kr, dkhong@lns.mit.edu}
\affiliation{Department of Physics, Pusan National University,
             Pusan 609-735, Korea}
\affiliation{Center for Theoretical Physics, Massachusetts
             Institute of Technology, Cambridge, MA 02139, USA}

\begin{abstract}
The nontrivial boundary conditions (BC's) for the Topcolor breaking
are investigated in the context of the TeV-scale extra dimension scenario.
In the gauge symmetry breaking mechanism via the BC's
we do not need to incorporate a dynamical mechanism for 
the Topcolor breaking into the model. 
Moreover, the Topcolor breaking can be realized
without introducing explicitly a (composite) scalar field.
We present a six dimensional model where the top and bottom quarks
in the bulk have the Topcolor charge while the other quarks in the bulk
do not. We also put the electroweak gauge interaction in
the six dimensional bulk.
The bottom quark condensation is naturally suppressed owing to
the power-like running of the bulk $U(1)_Y$ interaction, so that
only the top condensation is expected to take place.
We explore such a possibility based on the ladder Schwinger-Dyson equation
and show the cutoff to make the model viable.
\end{abstract}

\maketitle

\section{Introduction}

Recently, the model building along with the TeV-scale extra dimension
scenario~\cite{Antoniadis:1990ew,Dienes:1998vh} has been widely surveyed.
The gauge theories with extra dimensions have remarkable features.
Since the number of the Kaluza-Klein (KK) modes which is
the source of the attractive force increases explosively in high-energy,
the bulk gauge couplings grow very quickly and get strong.
Therefore the bulk gauge theories can naturally trigger
the dynamical chiral symmetry breaking (D$\chi$SB).~\cite{Dobrescu:1998dg, Cheng:1999bg, Arkani-Hamed:2000hv, Hashimoto:2000uk, Rius:2001dd, Abe:2001yi}

The top quark condensate~\cite{MTY89,Nambu89,Marciano89,Bardeen:1989ds},
which is often called the ``top mode standard model'' (TMSM), has been also
reexamined in the context of extra dimensions~\cite{Dobrescu:1998dg,Cheng:1999bg,Kobakhidze:1999ce,Arkani-Hamed:2000hv,Hashimoto:2000uk,Gusynin:2002cu,Hashimoto:2003ve,Gusynin:2004jp}.
In particular, Arkani-Hamed, Cheng, Dobrescu and
Hall (ACDH)~\cite{Arkani-Hamed:2000hv} proposed a version of the TMSM
where the third generation quarks and leptons as well as the
the Standard Model (SM) gauge bosons are put in the bulk,
while any four-fermion interactions are not introduced in the bulk
unlike the original version of the TMSM in four dimensions.
In Refs.~\cite{Hashimoto:2000uk,Gusynin:2002cu},
the full bulk gauge dynamics was investigated,
based on the ladder Schwinger-Dyson (SD) equation.
The phenomenological implications
were studied in Ref.~\cite{Hashimoto:2003ve}.
It is found that the model with $D=8$ can be viable and
both masses of the top quark and Higgs boson are
predicted as $m_t=172-175$ GeV and $m_H=176-188$ GeV, respectively.
However it turns out that
{\it the simplest scenario with $D=6$ does not work.}

On the other hand, it is known that 
field theories in six dimensions have several interesting features
relating to proton stability~\cite{Appelquist:2001mj},
explanation of the number of the generations of 
fermions~\cite{Dobrescu:2001ae}, etc..
In order to construct a viable top-condensate model in six dimensions,
we may introduce the four-fermion interaction in the bulk.
In other words, one of possibilities is the model building based on
the gauged Nambu-Jona-Lasinio (NJL) model
which is defined as the gauge theory with four-fermion interactions.
The phase structure of such a  gauged NJL model in the bulk was analyzed
in Ref.~\cite{Gusynin:2004jp}.
Another possibility is to introduce a strong gauge interaction such as
Topcolor in the bulk.
Topcolor models in four dimensions have been extensively
studied.~\cite{Hill:1991at,TC2,Dobrescu:1997nm}
(See for reviews Refs.~\cite{Cvetic:1997eb,Hill:2002ap}. )
The top seesaw mechanism can be realized by introducing five dimensional
right-handed top quark.~\cite{Cheng:1999bg, Cheng:2001nh, He:2001fz}

The Topcolor should be broken down in low energy.
In four dimensions, however, it is required to introduce some involved 
dynamical mechanism in order to break the Topcolor,
unless a (composite) scalar field is introduced for simplicity.
As for the gauge symmetry breaking,
the extra dimension scenario has an advantage as well. 
It is known that the gauge symmetry breaking can be easily achieved
in extra dimensions by imposing appropriate boundary conditions 
(BC's).~\cite{Kawamura:1999nj}
On the basis of more general BC's,
the Higgsless theory was proposed~\cite{Csaki:2003dt,Csaki:2003zu}
and has been investigated by a number of authors~\cite{Nomura:2003du,Barbieri:2003pr,Davoudiasl:2003me,Foadi:2003xa,Hirn:2004ze,Chivukula:2004kg,Davoudiasl:2004pw,Gabriel:2004ua,Evans:2004rc,Hewett:2004dv,Papucci:2004ip,Georgi:2004iy,Perelstein:2004sc}.
Although it may be difficult to construct Higgsless models consistent with 
the precision measurements, the gauge symmetry breaking mechanism via 
nontrivial BC's can be also applied to other models for the dynamical 
electroweak symmetry breaking. 
Such an attempt has not yet been discussed so far. 

A Topcolor model with nontrivial BC's obviously has some advantages:
We do not need to incorporate a dynamical mechanism for 
the Topcolor breaking.
We can break spontaneously the Topcolor gauge symmetry without 
introducing {\it explicitly} a (composite) scalar field.
We note here, however, that such a model has implicitly a scalar field
on the boundary.
In a sense, a theory with nontrivial BC's is equivalent to 
a model having a scalar field with a large vacuum expectation value (VEV)
which is put on the boundary.
In the infinite limit of the VEV the scalar field is completely decoupled
and hence the KK masses of the gauge boson are controlled only by
the compactification scale.
Therefore, we can neglect thoroughly such a scalar field.
In passing, the Topcolor gauge bosons do not have mass terms in the bulk
in the gauge breaking mechanism via the BC's.
The theory thus does not provide four-fermion (NJL-type) interactions 
in the bulk, unlike four dimensional Topcolor models.

In this paper we investigate the Topcolor breaking via nontrivial BC's
in six dimensions.
We assign the Topcolor charge, $SU(3)_1$, to the top and bottom quarks
in the bulk.
The quarks of the first and second generations in the bulk are assumed
to have the $SU(3)_2$ charge.
We then impose the nontrivial BC's so that
$SU(3)_1 \times SU(3)_2$ breaks down to the diagonal subgroup,
which is identified to QCD.
We also put the electroweak (EW) gauge interaction in the bulk.
The EW gauge sector is the same as the universal extra dimension
model~\cite{Appelquist:2000nn} with the composite Higgs field.
In order to obtain the chiral fermion in four dimensions,
we apply the compactification on a square proposed by Dobrescu and
Pont\'{o}n~\cite{Dobrescu:2004zi}, which is closely related to
the compactification on the orbifold $T^2/Z_4$.

For a viable model it is required that only the top condensation occurs
while other condensations such as bottom and leptons do not.
We call the requirement ``tMAC condition'' and the energy scale
``tMAC scale'' as in Ref.~\cite{Hashimoto:2003ve}.
Once we specify the model, the renormalization group (RG) flows of
the gauge couplings can be determined through the truncated KK effective
theory~\cite{Dienes:1998vh}. The running effects are very important to
study the tMAC scale.
We here note that the bulk hypercharge interaction $U(1)_Y$
rapidly becomes strong owing to the power-like running.
Thus the $U(1)$ tilting mechanism to suppress the bottom quark condensation
is automatically incorporated in the model.
The difference of the gauge coupling strengths between $SU(3)_1$ and 
$SU(3)_2$ leads to suppression of the up-, and charm-condensations.
When the theory behaves as a walking gauge
theory~\cite{Holdom:1984sk,Yamawaki:1985zg,Bando:1987we,Akiba:rr,AKW86}
and the gauge coupling of $SU(3)_1$ is very close to the critical coupling 
for the D$\chi$SB,
the situation that only the top quark condenses is naturally realized.
We analyze the tMAC scale by using the ladder SD equation
and depict the results in two dimensional plane of the cutoff $\Lambda$
and the ratio of the Topcolor and QCD couplings $g^2(R^{-1})/g_3^2(R^{-1})$
at the compactification scale $R^{-1} (\approx 10\;{\rm TeV})$.
For a slice $g^2(R^{-1})/g_3^2(R^{-1})=4.6$, for example,
we find that the tMAC scale is $\Lambda R \sim 10-10.5$.
We also show that the model is not excluded by constraints of 
S, T-parameters.

The paper is organized as follows:
In Sec.~\ref{sec2} we study the BC's for the Topcolor breaking.
In Sec.~\ref{sec3} we give the model and study running
effects of gauge couplings.
In Sec.~\ref{sec4} we determine the tMAC scale by solving
the ladder SD equation.
Sec.~\ref{summary} is devoted to summary and discussions.
In Appendix~\ref{app-a}, we give the chiral compactification on the square.
In Appendix~\ref{app-b}, we describe the condition that
the five dimensional gauge symmetry is restored on the boundaries.

\section{Boundary conditions for Topcolor breaking}
\label{sec2}

We explore possibility of the top quark condensate in six dimensions.
For simplicity, we compactify extra two spatial dimensions $(y^5,y^6)$
on a square with $0 \leq y^5, y^6 \leq L$.
We introduce the bulk Topcolor gauge interaction in order to
trigger the top condensation.
The Topcolor should be broken down in the low-energy effective theory.
In this section, we study nontrivial BC's for the Topcolor breaking.
First, we derive the BC's for the pure gauge theory in the bulk.
Next, we incorporate the top quark in the bulk.

\subsection{Pure gauge theory on a square}

Let us analyze the $SU(3)_1 \times SU(3)_2$ gauge theory in the bulk.
We assign the Topcolor to the $SU(3)_1$ gauge interaction.
The action is given by
\begin{equation}
S = \int d^4x \int_0^L dy^5 \int_0^L dy^6 {\cal L}_g,
\label{S}
\end{equation}
with the Lagrangian
\begin{equation}
  {\cal L}_g =
  - \frac{1}{4}F_{MN}^a F^{a\,MN} - \frac{1}{4}F_{MN}^{'a} F^{'a\,MN},
\end{equation}
where $M=0,1,2,3,5,6$, and
\begin{equation}
  F_{MN}^a \equiv \partial_M A_N^a - \partial_N A_M^a
  + g_{6D}^{} f^{abc} A_M^b A_N^c. \label{F}
\end{equation}
$f^{abc}$ is the structure constant of the gauge group,
$g_{6D}^{}$ the {\it dimensionful} bulk gauge coupling constant.
The definition of $F_{MN}^{'a}$ is the same as Eq.~(\ref{F})
with $A_M^{'a}$ and $g'_{6D}$.
The gauge fields $A_M^a$ and $A_M^{'a}$ are associated with the gauge groups
$SU(3)_1$ and $SU(3)_2$, respectively.
We also use the notation
\begin{equation}
  A_M \equiv A_M^a T^a,
\end{equation}
with $T^a$ being the generator of the SU(3) Lie algebra.

We break the gauge symmetry $SU(3)_1 \times SU(3)_2$ to the diagonal subgroup
by assigning nontrivial BC's to the gauge fields.
The unbroken subgroup is identified to the conventional QCD.

After integration by parts the variation of the action (\ref{S}) yields
\begin{eqnarray}
\lefteqn{
\delta S =
} \nonumber \\
&& \hspace*{-3mm}
 \int \! d^4x \!\! \int \! dy^5 dy^6 \left\{ \phantom{\bigg|}\!\!\!
 \left[\,
 \partial_M F^{aMN} - g_{6D}^{} f^{abc}F^{bMN}A_M^c\,\right]
 \delta\! A_N^{a} \right.\nonumber \\
&& \left. \hspace*{1.1cm}
+ \left[\,
 \partial_M F^{'aMN} - g'_{6D} f^{abc}F^{'bMN}A_M^{'c}\,\right]
 \delta\! A_N^{'a} \right\}\nonumber \\
&&  \hspace*{-3mm}
+\int \! d^4x \!\! \int \! dy^6
 \left(\,
 F_{5\mu}^a\,\delta\! A^{a\,\mu} + F_{5\mu}^{'a}\, \delta\! A^{'a\,\mu}
 \,\right)\bigg|_{(0,y^6)}^{(L,y^6)}
 \nonumber \\[2mm]
&&  \hspace*{-3mm}
+\int \! d^4x \!\! \int \! dy^5
 \left(\,
 F_{6\mu}^a\, \delta\! A^{a\,\mu} + F_{6\mu}^{'a}\, \delta\! A^{'a\,\mu}
 \,\right)\bigg|_{(y^5,0)}^{(y^5,L)}
 \nonumber \\[2mm]
&&  \hspace*{-3mm}
=0,
 \label{dS}
\end{eqnarray}
where
\begin{equation}
  X \bigg|_{(0,y)}^{(L,y)} \equiv X(x^\mu,L,y)-X(x^\mu,0,y) , \label{X}
\end{equation}
and similar is the definition of $X |_{(y,0)}^{(y,L)}$.
The vanishing requirement of the first term in Eq.~(\ref{dS}) corresponds
to the equation of motion.

The zero modes of the gauge scalar fields $A_{5,6}^{(')}$ should be
eliminated from the low-energy spectrum.
We thus impose the Dirichlet BC's on the gauge scalars,
\begin{equation}
 \left\{ \begin{array}{ll}
  A_5 (0,y) = A_5 (L,y) = 0, \quad A'_5 (0,y) = A'_5 (L,y) = 0, \\[2mm]
  A_6 (y,0) = A_6 (y,L) = 0, \quad A'_6 (y,0) = A'_6 (y,L) = 0,
 \end{array}\right.
 \label{bc-gs}
\end{equation}
where we abbreviated the trivial argument $x^\mu$ in
$A_5(x^\mu,0,y)$, etc.
We rewrite the two integrals $\int dy^6, \int dy^5$ in Eq.~(\ref{dS})
to $\int dy$ by introducing a single dummy index $y$.
Then we obtain
a BC for $A_\mu^a$ and $A_\mu^{'a}$,
\begin{eqnarray}
&&  \phantom{+}  \left(\, \partial_5 A_\mu^a\, \delta\! A^{a\,\mu}
 + \partial_5 A_\mu^{'a}\, \delta\! A^{'a\,\mu}
 \,\right)\bigg|_{(0,y)}^{(L,y)}
 \nonumber \\[2mm]
&&+
   \left(\, \partial_6 A_\mu^a\, \delta\! A^{a\,\mu}
 + \partial_6 A_\mu^{'a}\, \delta\! A^{'a\,\mu}
 \,\right)\bigg|_{(y,0)}^{(y,L)}=0.
 \label{bc}
\end{eqnarray}
If the variations $\delta A_\mu^{(')}$ on the boundaries,
$(0\,{\rm or}\,L,y)$ and $(y,0\,{\rm or}\,L)$, are independent,
Eq.~(\ref{bc}) yields two BC's,
\begin{equation}
 \left(\,
   \partial_5 A_\mu^a \,\delta\! A^{a\,\mu}
 + \partial_5 A_\mu^{'a}\, \delta\! A^{'a\,\mu}
 \,\right)\bigg|_{(0,y)}^{(L,y)}
 = 0,
 \label{bc-1}
\end{equation}
and
\begin{equation}
 \left(\,
  \partial_6 A_\mu^a \,\delta\! A^{a\,\mu}
 + \partial_6 A_\mu^{'a}\, \delta\! A^{'a\,\mu}
 \,\right)\bigg|_{(y,0)}^{(y,L)}
 = 0.
 \label{bc-2}
\end{equation}
Since we adopt later on a chiral compactification on the square
with two adjacent sides identified,
we use a general expression (\ref{bc}) in the following discussion.

Now we further assign the following BC to $A_\mu$ and $A'_\mu$
on every boundary,
\begin{equation}
 A_\mu |^{(0,y),(L,y),(y,0),(y,L)} =
 \tan\theta \,A'_\mu|^{(0,y),(L,y),(y,0),(y,L)},
 \label{bc-A}
\end{equation}
where $\theta$ is a constant.
Note that the derivative terms are not identified at the boundary, i.e.,
$\partial_5 A_\mu |^{(0,y)} \ne \tan\theta\,\partial_5 A'_\mu |^{(0,y)}$, etc.
The BC (\ref{bc}) is then rewritten as
\begin{eqnarray}
&& \phantom{+} \partial_5 \left[\,A_\mu^{'a} + \tan \theta A_\mu^a\right]
   \delta A^{'a\,\mu}\bigg|_{(0,y)}^{(L,y)}
   \nonumber \\[2mm]
&& + \partial_6 \left[\,A_\mu^{'a} + \tan \theta A_\mu^a\right]
   \delta A^{'a\,\mu}\bigg|_{(y,0)}^{(y,L)}=0 .
 \label{bc-3}
\end{eqnarray}
We here define the ``gluon'' field $G_\mu$ and
the ``coloron'' field $G'_\mu$ as
\begin{equation}
 \left\{
  \begin{array}{lcl}
  G_\mu(x^\mu,y^5,y^6) & = & \phantom{+}
  A'_\mu \cos\theta + A_\mu \sin\theta , \\[2mm]
  G'_\mu(x^\mu,y^5,y^6) & = &
  - A'_\mu \sin\theta + A_\mu \cos\theta .
  \end{array}
 \right. \label{G-A}
\end{equation}
The gluon field should have zero modes.
We thus impose the Neumann BC's on the gluon field $G_\mu$:
\begin{equation}
 \partial_5 G_\mu^a |^{(0,y),(L,y)} = 0, \qquad
 \partial_6 G_\mu^a |^{(y,0),(y,L)} = 0.
 \label{bc-G0}
\end{equation}
By definition (\ref{G-A}),
Eq.~(\ref{bc-A}) automatically
yields the Dirichlet BC's for $G'_\mu$:
\begin{equation}
 G'_\mu(0,y) = G'_\mu (L,y) = G'_\mu(y,0) = G'_\mu (y,L) = 0.
 \label{bc-Gp0}
\end{equation}
Hence we obtain the KK decompositions for $G_\mu$ and $G'_\mu$,
\begin{eqnarray}
  G_\mu(x^\mu,y^5,y^6) &=&
  \frac{1}{L} \sum_{n_5,n_6 \geq 0} G_{\mu}^{[n_5, n_6]}(x^\mu)
  \nonumber \\ && \hspace*{-1.5cm} \times {\cal N}
  \left[\,
  \cos\left(\dfrac{\pi}{L} n_5 y^5\right)
  \cos\left(\dfrac{\pi}{L} n_6 y^6\right)\,\right],
\end{eqnarray}
with
\begin{equation}
 {\cal N} \equiv 2\sqrt{\frac{1}{(1+\delta_{n_5,0})(1+\delta_{n_6,0})}},
\end{equation}
and
\begin{eqnarray}
  G'_\mu(x^\mu,y^5,y^6) &=&
  \frac{1}{L} \sum_{n_5, n_6 > 0} G_{\mu}^{'\,[n_5, n_6]}(x^\mu)
  \nonumber \\ && \hspace*{-1.5cm} \times 2
  \left[\,
  \sin\left(\dfrac{\pi}{L} n_5 y^5\right)
  \sin\left(\dfrac{\pi}{L} n_6 y^6\right)\,\right],
\end{eqnarray}
respectively.
We here note that only $G_\mu$ associated with the diagonal subgroup
includes a zero mode, while $G'_\mu$ does not.
We identify the unbroken gauge group to QCD, $SU(3)_c$.

We comment on our choice of the BC's.
Under the above identification the gauge symmetry breaking
$SU(3)_1 \times SU(3)_2 \to SU(3)_c$ takes place
on all the boundaries.
This is not a unique choice: for example, we can also construct a model
in which the gauge symmetry is broken down only at a part of boundary
like $(0,y),(L,y)$.
We choose the BC's (\ref{bc-G0})--(\ref{bc-Gp0}) to be consistent with
the chiral compactification.

\subsection{Topcolor model on a square}

Let us take into account the top quark $T$ in the bulk,
which has the $SU(3)_1$ charge,
\begin{equation}
  {\cal L}_t = \bar{T}_+\,i D_M \Gamma^M T_+
 + \bar{T}_- \,i D_M \Gamma^M T_-,
\end{equation}
with
\begin{equation}
  D_M \equiv \left(\frac{1}{2}\DLR_M - i g_{6D}^{} A_M\right) ,
\end{equation}
where
\begin{equation}
  \bar{T}\!\DLR_M \!\!\Gamma^M T \equiv
  \bar{T} \Gamma^M (\partial_M T)  -   (\partial_M \bar{T}) \Gamma^M T ,
\end{equation}
and $\Gamma^M$'s are the Gamma matrices in six dimensions.
The chiral fermions $T_\pm$ in the bulk are defined by
\begin{equation}
  T_\pm \equiv \frac{1}{2}(1 \pm \Gamma_{\chi,7})\,T,
\end{equation}
where $\Gamma_{\chi,7}$ is the chirality matrix in six dimensions.
The theory obviously has the chiral symmetry.
We here note that the chiral fermions $T_\pm$ contain both of
the right and left handed components, i.e.,
\begin{equation}
  T_\pm = T_{\pm R} + T_{\pm L}, \quad
  T_{\pm R,L} \equiv \frac{1}{2}(1 \pm \Gamma_{\chi,5})\,T_\pm,
\end{equation}
with the four dimensional chirality matrix $\Gamma_{\chi,5}$.
Therefore we must carry out the chiral compactification
in order to obtain the SM-like top quark in low-energy.

Following Dobrescu and Pont\'{o}n~\cite{Dobrescu:2004zi},
we identify two adjacent sides as follows:
\begin{equation}
  (y,0) \equiv (0,y), \quad (y,L) \equiv (L,y), \quad
  {}^\forall \!y \in [0,L] ,
\end{equation}
which is closely related to the orbifold compactification on $T^2/Z_4$.
We take a notation that $T_{+R,-L}$ include
the SM-like top quarks $t_{R,L}$ as the zero modes.
In general, the value of a field at two identified points differs by a
nontrivial phase, if a loop around the point is noncontractible.
As in \cite{Dobrescu:2004zi}, we assign the following BC's
\begin{equation}
  T_{+R}(y,0) = T_{+R}(0,y), \quad
  T_{+R}(y,L) = T_{+R}(L,y),
\end{equation}
to the fermion $T_{+R}$.
The BC's for $T_{-L}$ are the same.
For $T_{+L}$ and $T_{-R}$ different BC's should be imposed.
For details, see Appendix~\ref{app-a} and Ref.~\cite{Dobrescu:2004zi}.
On the other hand, for gauge fields $G_\mu$ and $G'_\mu$
the chiral compactification further requires
\begin{equation}
 G_\mu(0,y) = G_\mu (y,0) , \qquad G_\mu(y,L) = G_\mu (L,y),
 \label{bc-G}
\end{equation}
and
\begin{equation}
 \partial_6 G'_\mu |^{(y,0)} = -\partial_5 G'_\mu |^{(0,y)},
 \quad
 \partial_6 G'_\mu |^{(y,L)} = -\partial_5 G'_\mu |^{(L,y)},
 \label{bc-Gp}
\end{equation}
in addition to the BC's (\ref{bc-G0})--(\ref{bc-Gp0}).
It is natural to require that on the boundaries
the theory is reduced into the five dimensional one.
Details are summarized in Appendix~\ref{app-b}.
We then find that the desirable BC's for the derivative terms of $T$ are
\begin{equation}
 \partial_5 T_{+R,-L}|^{(0,y),(L,y)} = 0 , \quad
 \partial_6 T_{+R,-L}|^{(y,0),(y,L)} = 0 .
 \label{bc-dT}
\end{equation}
We here note that Eqs.~(\ref{bc-gs}) also imply Eq.~(\ref{bc-dT}).
The KK decompositions of $T_{+R,-L}$, $G_\mu$ and $G'_\mu$ are obtained as
\begin{equation}
  T_{+R,-L}(x^\mu,y^5,y^6) =
  \frac{1}{L} \sum_{j \geq k \geq 0} T_{+R,-L}^{[j, k]}(x^\mu)
  f_{cc}^{[j,k]}(y^5,y^6) ,
\label{tR}
\end{equation}
\begin{equation}
  G_\mu(x^\mu,y^5,y^6) =
  \frac{1}{L} \sum_{j \geq k \geq 0} G_{\mu}^{[j, k]}(x^\mu)
  f_{cc}^{[j,k]}(y^5,y^6), \label{Gkk}
\end{equation}
\begin{equation}
  G'_\mu(x^\mu,y^5,y^6) =
  \frac{1}{L} \sum_{j > k > 0} G_{\mu}^{'\,[j, k]}(x^\mu)
  f_{ss}^{[j,k]}(y^5,y^6), \label{Gpkk}
\end{equation}
with
\begin{eqnarray}
  f_{cc}^{[j,k]} &\equiv& {\cal N}_{cc}
  \left[\,
  \cos\left(\dfrac{\pi}{L} j y^5\right)
  \cos\left(\dfrac{\pi}{L} k y^6\right)
  \right.  \nonumber \\ && \qquad \left.
 +\cos\left(\dfrac{\pi}{L} k y^5\right)
  \cos\left(\dfrac{\pi}{L} j y^6\right) \,\right],
\end{eqnarray}
\begin{eqnarray}
  f_{ss}^{[j,k]} &\equiv& {\cal N}_{ss}
  \left[\,
  \sin\left(\dfrac{\pi}{L} j y^5\right)
  \sin\left(\dfrac{\pi}{L} k y^6\right)
  \right.  \nonumber \\ && \qquad \left.
 -\sin\left(\dfrac{\pi}{L} k y^5\right)
  \sin\left(\dfrac{\pi}{L} j y^6\right) \,\right] ,
\end{eqnarray}
where ${\cal N}_{cc}$ and ${\cal N}_{ss}$ are the normalization factors
given in Appendix~\ref{app-a}.
In particular, the function $f_{cc}^{[0,0]}$ for the zero mode is given by
\begin{equation}
  f_{cc}^{[0,0]} = 1.
\end{equation}

In our compactification, the KK masses for $G_\mu^{[j,k]}$ and
$G_\mu^{'[j,k]}$ are labeled by integers $j$ and $k$ as
\begin{equation}
  M_{j,k}^2 \equiv \frac{\pi^2}{L^2} (j^2+k^2).
\end{equation}
The lightest KK mass for $G_\mu$ is, as usual,
\begin{equation}
  M_G \equiv M_{1,0} =  \frac{\pi}{L}.
\end{equation}
However, the coloron field does not include the KK components of
$(j > 0,k=0)$ and $(j,k=j)$.
Therefore the lowest KK mass for the coloron is given by
\begin{equation}
  M_C \equiv M_{2,1} =  \frac{\sqrt{5}\pi}{L}.
\end{equation}
We also comment that the total number of KK modes for $G'_\mu$ below
the cutoff $\Lambda$ is slightly smaller than that for $G_\mu$.
Such a difference is, however, negligible for a large $\Lambda$.

{}From the symmetry breaking pattern $SU(3)_1 \times SU(3)_2 \to SU(3)_c$,
the gauge couplings of $SU(3)_1$ and $SU(3)_2$ are not arbitrary,
but they are related to the QCD coupling.
Integrating the six dimensional Lagrangian over $dy^5$ and $dy^6$,
we define the four dimensional theory,
\begin{equation}
  {\cal L}_{4D} \equiv \int_0^L dy^5 \int_0^L dy^6 {\cal L}_{6D} ,
  \label{4D-6D}
\end{equation}
with
\begin{equation}
  {\cal L}_{6D} = {\cal L}_t + {\cal L}_g.
\end{equation}
By using Eqs.~(\ref{tR})--(\ref{Gpkk}) and the definition (\ref{G-A}),
we find the interaction term between zero modes of the top and the gluon as
\begin{equation}
  {\cal L}_{\rm int} = \frac{g_{6D}\sin\theta}{L}
  \bar{T}_{+R,-L}^{[0,0]}\Gamma^\mu G_\mu^{[0,0]}
  T_{+R,-L}^{[0,0]} . \label{L-int}
\end{equation}
We here note that the definition (\ref{4D-6D}) implies the relations
between the six and four dimensional gauge couplings as
\begin{equation}
  g_{6D}^2 = L^2 g^2, \quad  g_{6D}^{'2} = L^2 g'{}^2, \label{gd}
\end{equation}
where $g$ and $g'$ denote the four dimensional gauge coupling constants
for $SU(3)_1$ and $SU(3)_2$, respectively.
Eq.~(\ref{L-int}) then yields the relation
\begin{equation}
  g_3 = g \sin\theta, \label{g-g3}
\end{equation}
where $g_3$ is the four dimensional QCD coupling.
In the same way, we obtain a similar relation between QCD and
$SU(3)_2$ couplings,
\begin{equation}
  g_3 = g' \cos\theta. \label{gp-g3}
\end{equation}
Eqs.~(\ref{g-g3})--(\ref{gp-g3}) read
\begin{equation}
  \frac{1}{g_3^2} = \frac{1}{g^2}+\frac{1}{g^{'2}}.
  \label{g3-g}
\end{equation}

\section{The model}
\label{sec3}

\begin{table}[tb]
 \centering
  \begin{tabular}{|c|cccc|}\hline
   & $SU(3)_1$ & $SU(3)_2$ & $SU(2)_W$ & $U(1)_Y$ \\ \hline
   $(t,b)_-$ & \bf{3} & \bf{1} & \bf{2} & $1/6$ \\
   $t_+$ & \bf{3} & \bf{1} & \bf{1} & $2/3$ \\
   $b_+$ & \bf{3} & \bf{1} & \bf{1} & $-1/3$ \\
   $(\nu_\tau,\tau)_-$ & \bf{1} & \bf{1} & \bf{2} & $-1/2$ \\
   $\tau_+$ & \bf{1} & \bf{1} & \bf{1} & $-1$ \\ \hline \hline
   $(c,s)_-$ & \bf{1} & \bf{3} & \bf{2} & $1/6$ \\
   $c_+$ & \bf{1} & \bf{3} & \bf{1} & $2/3$ \\
   $s_+$ & \bf{1} & \bf{3} & \bf{1} & $-1/3$ \\
   $(\nu_\mu,\mu)_-$ & \bf{1} & \bf{1} & \bf{2} & $-1/2$ \\
   $\mu_+$ & \bf{1} & \bf{1} & \bf{1} & $-1$ \\ \hline \hline
   $(u,d)_-$ & \bf{1} & \bf{3} & \bf{2} & $1/6$ \\
   $u_+$ & \bf{1} & \bf{3} & \bf{1} & $2/3$ \\
   $d_+$ & \bf{1} & \bf{3} & \bf{1} & $-1/3$ \\
   $(\nu_e,e)_-$ & \bf{1} & \bf{1} & \bf{2} & $-1/2$ \\
   $e_+$ & \bf{1} & \bf{1} & \bf{1} & $-1$ \\ \hline \hline
   $\psi_X$ & \bf{3} & \bf{1} & \bf{1} & $0$ \\ \hline
  \end{tabular}
  \caption{The charge assignment of the model. \label{tab1}}
\end{table}

We now incorporate all quarks and leptons of the SM into the model.
We put all of gauge fields and SM fermions in the six dimensional bulk.
We may introduce right-handed neutrinos in the bulk, which is not
relevant in the following analysis.

Let us assign the bulk top and bottom quarks to
the $SU(3)_1$ charge while the quarks of the first and second generations
to the $SU(3)_2$ charge.
We assume that the electroweak gauge sector is the same as the model of
the universal extra dimensions~\cite{Appelquist:2000nn}.
We perform the chiral compactification described in Sec.~\ref{sec2},
Appendix~\ref{app-a}, and Ref.~\cite{Dobrescu:2004zi}.
The Topcolor interaction should be sufficiently strong to trigger
the top condensation, so that
we may further introduce vector-like (heavy) fermions $\psi_X$
having the $SU(3)_1$ charge in order to adjust the RG flow of $SU(3)_1$.
We show the charge assignment in Table~\ref{tab1}.

While $SU(3)_1$ and $SU(3)_2$ are vector-like,
the $SU(2)_W$ and $U(1)_Y$ representations are chiral.
Although the six dimensional theory is anomalous under
the charge assignment in Table~\ref{tab1}, the anomalies can be cancelled
out by the Green-Schwarz mechanism~\cite{Green:sg}.
We assume that the Green-Schwarz counterterm does not change
the results in the following analysis.

Let us study running of gauge couplings in
the ``truncated KK'' effective theory~\cite{Dienes:1998vh}
based on the $\overline{\rm MS}$-scheme.
In this section, we use the unit of the extra momentum $R^{-1}$
instead of $L$,
\begin{equation}
  R^{-1} \equiv \frac{\pi}{L} .
\end{equation}
We expand bulk fields into KK modes and
construct a four dimensional effective theory.
Below $R^{-1}$ the renormalization group equations (RGEs) 
of the four dimensional gauge couplings
$g_i (i=3,2,Y)$ are given by those of the SM,
\begin{equation}
  (4\pi)^2 \mu \frac{d g_i}{d \mu} = b_i\,g_i^3, \quad (\mu < R^{-1})
\end{equation}
with $b_3=-7, b_2=-\frac{19}{6}$ and $b_Y=\frac{41}{6}$.
Above $R^{-1}$
QCD should be replaced by
the $SU(3)_1 \times SU(3)_2$ gauge interaction.
We also need to take into account contributions
of KK modes in $\mu \geq R^{-1}$.
Since the KK modes heavier than the renormalization scale $\mu$
are decoupled in the $\overline{\rm MS}$-RGEs,
we only need summing up the loops of the KK modes lighter than $\mu$.
We estimate the total number of KK modes below $\mu$ by the volume
of the momentum space of extra dimensions dividing by the identification
factor $n$,
\begin{equation}
N_{\rm KK}(\mu) = \frac{\pi (\mu R)^2}{n} , \quad (\mu \gg R^{-1}) .
  \label{nkk_app}
\end{equation}
Note that we impose additional BC's such as Eq.~(\ref{bc-dT})
other than the BC's for the $T^2/Z_4$ compactification.
Therefore our model corresponds to the case of 
\begin{equation}
  n=8.
\end{equation}
The estimate (\ref{nkk_app}) works well for $\mu R \gg 1$.
(See, e.g. Ref.~\cite{Hashimoto:2003ve}.~)
Within the truncated KK effective theory, we obtain the RGE
\begin{equation}
  (4\pi)^2 \mu \frac{d g}{d \mu} =
  N_{\rm KK} (\mu) \,b_{\rm tc} \,g^3,
  \quad (\mu \geq R^{-1}) \label{rge_ED}
\end{equation}
with
\begin{equation}
  b_{\rm tc} = -\frac{22}{3} \, + \frac{4}{3} \cdot N_X, \quad
 \mbox{for} \quad SU(3)_1, \label{b3KK}
\end{equation}
where $N_X$ is the number of $\psi_X$ with the fundamental representation.
Other RGE coefficients are given by
\begin{align}
  b' &= -\frac{14}{3} , & {\rm for} & \quad SU(3)_2, \\
  b'_2 &= \frac{4}{3} + \frac{1}{6} \, n_h,  & {\rm for} & \quad SU(2)_W, \\
  b'_Y &= \frac{40}{3} + \frac{1}{6} \, n_h, & {\rm for} & \quad U(1)_Y.
\end{align}
In the following analysis, we assume that one composite Higgs doublet
appears in the low-energy spectrum, i.e., $n_h=1$.

When the RG flow of $SU(3)_1$ is ``walking'' more slowly than that of
$SU(3)_2$, the top condensation is favored rather than the up and charm
condensations.
Thus we require $b_{\rm tc} > b'$, i.e.,
\begin{equation}
  N_X \geq 3 .
\end{equation}

We now define the {\it dimensionless} bulk gauge coupling $\hat g$ as
$\hat g^2 \equiv g_{6D}^2 \mu^2$
and thereby obtain
\begin{equation}
  \hat g^2(\mu) = (\pi R \mu)^2 g^2 (\mu),
  \label{hat-g}
\end{equation}
where we used Eq.~(\ref{gd}).
Combining Eq.~(\ref{hat-g}) with the RGE~(\ref{rge_ED}),
we find RGEs for the dimensionless bulk Topcolor coupling $\hat g$,
\begin{equation}
 \mu \frac{d}{d \mu} \hat g = \hat g
 + \NDA\, b_{\rm tc}\, \hat g^3 , \label{rge_ED2}
\end{equation}
with $\NDA$ being the $D$-dimensional loop factor in 
the naive dimensional analysis (NDA),
\begin{equation}
  \NDA \equiv \frac{1}{(4\pi)^{D/2}\Gamma(D/2)} .
\end{equation}
The RGEs for $SU(3)_2$, $SU(2)_W$, and $U(1)_Y$ are the same as
Eq.~(\ref{rge_ED2}).

\begin{figure}[tbp]
  \begin{center}
    \resizebox{0.45\textwidth}{!}{
     \includegraphics{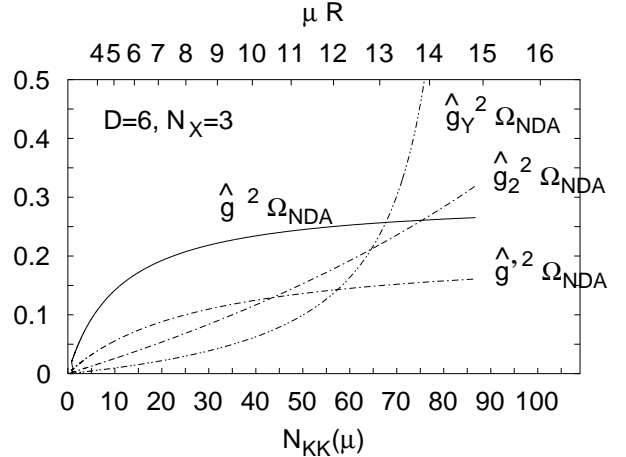}}
    \caption{Typical RG flows for the model with $N_X=3$.
             The ratio of the Topcolor and QCD coupling
             constants at $R^{-1}=$ 10 TeV is taken to
             $g^2(R^{-1})/g_3^2(R^{-1}) = 4.5$.
             \label{fig-hatg}}
  \end{center}
\end{figure}

Once we specify $N_X$ and the Topcolor coupling at $R^{-1}$,
the RG flow of $\hat g^2$ is completely determined.
(See also Eq.~(\ref{g3-g}). )
We show typical RG flows in Fig.~\ref{fig-hatg}.
We used the following values of $\alpha_i (\equiv g_i^2/(4\pi))$
at $\mu=M_Z(=91.1876 \;{\rm GeV})$
as inputs of RGEs:~\cite{PDG}
\begin{eqnarray}
  \alpha_3(M_Z) &=& 0.1172, \label{qcd-mz} \\
  \alpha_2(M_Z) &=& 0.033822, \label{su2-mz} \\
  \alpha_Y(M_Z) &=& 0.010167 . \label{u1-mz}
\end{eqnarray}
We also note the value of $\alpha_3$ at $R^{-1}=$ 10 TeV evolved by
the 1-loop RGE,
\begin{equation}
    \alpha_3(\mbox{10 TeV}) = 0.07264.
\end{equation}

The $U(1)_Y$ gauge interaction has the Landau pole $\Ly$
at which the gauge coupling diverges. (See Fig.~\ref{fig-hatg}. )
The bulk gauge coupling $\hat g_Y (\mu)$ rapidly grows
due to the power-like behavior of the running.
As a result, the Landau pole $\Ly$ is not so far from the compactification
scale $R^{-1}$.
We thus need to introduce a cutoff $\Lambda$ smaller than
the Landau pole $\Ly$.

\section{Analysis of the ladder SD equation}
\label{sec4}

We explore the energy region where
only the top quark condenses while others do not (tMAC region).
Since our model explicitly breaks the six dimensional Lorentz symmetry,
it is not obvious whether or not the approach of the ladder SD equation
for the bulk fermion is appropriate.
Nevertheless we may adopt the ladder SD equation in six dimensions,
supposing the cutoff $\Lambda R \sim {\cal O}(10)$ is large enough.

The power-like running of the gauge couplings is crucial for
the analysis of the tMAC region.
Thus we should incorporate the running effects in the ladder SD equation.
Several methods have been applied to the phenomenology of the low-energy
QCD in four dimensions.
Simplest one is the Higashijima-Miransky approximation
in which the gauge coupling is replaced by~\cite{imp-simplest}
\begin{equation}
  g^2 \to g^2 (\max(-p^2,-q^2)),
\end{equation}
where $p$ and $q$ are external and loop momenta of the fermion, respectively.
However the Higashijima-Miransky approximation is inconsistent with
the axial Ward-Takahashi (WT) identity.
A natural choice is to take the argument of $g$ to
the gluon loop momentum $(p-q)$,
\begin{equation}
  g^2 \to g^2 (-(p-q)^2) .
\end{equation}
In this case, the ladder approximation can be consistent with
both of vector and axial WT identities~\cite{Kugo:1992pr}.
A demerit of the method is that the angular integration cannot be performed
analytically, i.e., the numerical calculation becomes complicated.
In Ref.~\cite{Aoki:1990aq}, it is shown that the approximation
\begin{equation}
  g^2 \to g^2 (-(p^2+q^2)) \label{g-ave}
\end{equation}
works well in four dimensions.
We may adopt Eq.~(\ref{g-ave}) even in extra dimensions.

Let us solve the ladder SD equation including running effects.
For consistency with the vector Ward-Takahashi identity,
we choose the Landau gauge and then obtain the ladder
SD equation for the fermion mass function $B_f$ as follows:
\begin{eqnarray}
 B_f (x) &=& (D-1)\int_{R^{-2}}^{\Lambda^2} \!\!dy\,
 y^{D/2-1}\frac{B_f(y)}{y+B_f^2(y)}\frac{\kappa_f (x+y)}{x+y}
 \nonumber \\ && \qquad \qquad \times
 \Bigg[\,K_{B} (x,y) + (x \leftrightarrow y) \,\Bigg],
 \label{sd_b_imp}
\end{eqnarray}
with $f=t,b,c,u,\ell$, and $x \equiv -p^2$, and $y \equiv -q^2$, where
the kernel $K_B$ is given by~\cite{Hashimoto:2000uk}
\begin{equation}
  K_{B}(x,y) = \frac{1}{x}\left(1-\frac{y}{3x}\right) \theta(x-y), \quad
  \mbox{for } D=6.
\end{equation}
We identified the infrared (IR) cutoff of the SD equation to
the compactification scale $R^{-1}$.
The binding strengths $\kappa_f$'s are
\begin{align}
  \kappa_t (\mu^2)&= C_F \hat g^2 (\mu) \NDA
                 \,+\, \frac{1}{9}\hat g_Y^2 (\mu) \NDA \label{k_t}, \\
  \kappa_b (\mu^2)&= C_F \hat g^2 (\mu) \NDA
                 \,-\, \frac{1}{18}\hat g_Y^2 (\mu) \NDA \label{k_b}, \\
  \kappa_{c,u} (\mu^2)&= C_F \hat g'{}^2 (\mu) \NDA
                 \,+\, \frac{1}{9}\hat g_Y^2 (\mu) \NDA \label{k_c}, \\
  \kappa_\ell (\mu^2)&= \phantom{C_F \hat g^2 (\mu) \NDA + \;\;\;\;}
                   \frac{1}{2} \hat g_Y^2 (\mu) \NDA \label{k_l},
\end{align}
for the top, bottom, charm, up and lepton condensates, respectively.
The constant $C_F (= 4/3)$ is the quadratic Casimir of
the fundamental representation of $SU(3)$.
In the following analysis, we study these four channels.
The argument of $\kappa_f$ should be smaller than the Landau pole of
$U(1)_Y$, i.e.,
\begin{equation}
  \max(x+y) = 2\Lambda^2 < \Ly^2 .
\end{equation}

\begin{figure}[tbp]
  \begin{center}
    \resizebox{0.45\textwidth}{!}{
     \includegraphics{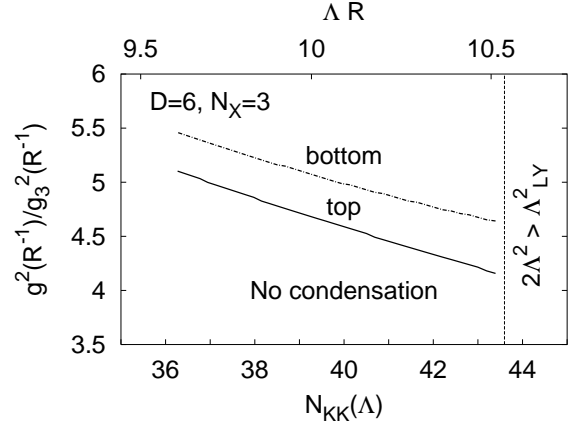}}
    \caption{The tMAC region for the model with $N_X=3$.
             The coupling constants of the Topcolor and QCD
             at the compactification scale $R^{-1}$ are represented as
             $g^2(R^{-1})$ and $g_3^2(R^{-1})$, respectively.
             In the ``top'' region only the top condensation occurs
             (tMAC region). In the ``bottom'' region
             the bottom quark condenses as well.
             No condensation takes place in the region of ``No condensation''.
             For $2\Lambda^2 > \Ly^2$ the argument of the gauge coupling
             in the kernel of the ladder SD equation exceeds the
             Landau pole of $U(1)_Y$.
             \label{fig1}}
  \end{center}
\end{figure}

We numerically solve the SD equation by using the iteration method, whose
details are described in Ref.~\cite{Hashimoto:2000uk}.
In the analysis, we fix the compactification scale $R^{-1}$ to 10 TeV.
For other values, the results are essentially unchanged.
We depict the result for the models with $N_X=3$ in Fig.~\ref{fig1}.
The ``top'' region in Fig.~\ref{fig1} corresponds to the tMAC.
If we choose the ratio of the values of the Topcolor and QCD couplings
at $R^{-1}=$ 10 TeV to
$g^2(R^{-1})/g_3^2(R^{-1}) \sim 4.2-4.6$,
the tMAC region is $\Lambda R \sim 10-10.5$.
In the region it turns out that the up- and
charm-condensations do not occur.
For $\Lambda R > 10.5$ the lepton condensation is favored.

Similarly, the tMAC regions are also found for models with $N_X=4,5$.
However the regions become narrower:
for $g^2(R^{-1})/g_3^2(R^{-1}) \sim 2.1-2.3$, $\Lambda R \sim 10.2-10.5$,
$(N_X=4)$;
for $g^2(R^{-1})/g_3^2(R^{-1}) \sim 1.3-1.4$, $\Lambda R \sim 10.3-10.5$,
$(N_X=5)$.

\section{Summary and discussions}
\label{summary}

We studied the Topcolor model in the six dimensional bulk.
We assigned the nontrivial BC's to the Topcolor gauge fields
so that the Topcolor is broken down on the boundaries.
As a three generation model
we considered the model whose charge assignments are shown in
Table~\ref{tab1}.
Since the top and bottom quarks have the Topcolor charge while
the other quarks do not in the model, the up and charm condensations
are unlikely to occur.
When the bulk $U(1)_Y$ interaction is sufficiently strong,
the bottom condensation is also suppressed.
In this way, we can expect that only the top quark condenses,
which is required for a viable model.
In order to demonstrate the existence of such a situation,
we analyzed the ladder SD equation including the RGE effects of
the bulk gauge couplings.
We then found that the situation can be realized in the ``top'' region
shown in Fig.~\ref{fig1}, which is the result for the model
with three extra (heavy) vector-like fermions having the Topcolor charge,
i.e., $N_X=3$.
For example, when the ratio of the couplings of Topcolor and QCD is
taken to $g^2(R^{-1})/g_3^2(R^{-1}) \sim 4.2-4.6$ with
$R^{-1}(\approx \mbox{10 TeV})$ being the compactification scale,
the cutoff $\Lambda$ should be $\Lambda R \sim 10-10.5$.
The models with $N_X=4,5$ may be possible as well.

The electroweak gauge sector of the model is the same as the universal
extra dimension model~\cite{Appelquist:2000nn}.
The compactification scale $R^{-1}$ is severely constrained by
the LEP precision data~\cite{PDG}.
Since the KK modes of bulk fermions are vector-like, the constraint from
the $S$ parameter is suppressed,
$S \approx 10^{-2} \sum_{j,k}\frac{m_t^2}{M_{j,k}^2}$.
Hence the $T$-parameter constraint is essential.
We may estimate the $T$-parameter as in Ref.~\cite{Appelquist:2000nn},
\begin{equation}
  T \approx 0.76 \sum_{j,k}\frac{m_t^2}{M_{j,k}^2} ,
\end{equation}
where we neglected ${\cal O}(m_t^4/M_{j,k}^4)$ contributions.
When we take $\max(M_{j,k})=\Lambda$ or $\sqrt{2}\Lambda$ with
$\Lambda \sim (10-10.5)R^{-1}$, the estimate of the $T$-parameter is
\begin{equation}
T \approx (4-5)\times 10^{-2}\frac{(1\;\mbox{TeV}^2)}{R^{-2}(\mbox{TeV}^2)}.
\end{equation}
The current constraint $T < 0.02$ at $95\%$ C.L. with the Higgs boson mass
$m_H=117$ GeV~\cite{PDG} yields $R^{-1} > 1.4-1.6$ TeV.
For larger $m_H$ the lower bound of $R^{-1}$ gets smaller.
For the reference value $R^{-1}=10$ TeV,
we can expect that the contributions of KK modes to the $T$-parameter is
negligibly small,
even if we take into account errors arising from nonperturbative effects.
In this case, however, we may need to allow the fine tuning of
${\cal O}(m_t^2 R^2) \sim 3 \times 10^{-4}$.

Our approach is very sensitive to the cutoff, i.e., the UV physics.
The UV completion by theory space~\cite{Arkani-Hamed:2001ca,Hill:2000mu}
may be required.

\section*{Acknowledgments}
We are thankful to J. Minahan and V. Miransky for useful comments.
The work is supported by KRF PBRG 2002-070-C00022 and also in part by
funds provided by the U.S. Department of Energy (D.O.E.)
under cooperative research agreement \#DF-FC02-94ER40818.

\appendix
\section{Square Compactification}
\label{app-a}

We summarize the chiral compactification on the square
with $0 \leq y^5, y^6 \leq L$.
Following Dobrescu and Pont\'{o}n~\cite{Dobrescu:2004zi},
we identify two adjacent sides as
\begin{equation}
  (y,0) \equiv (0,y), \quad (y,L) \equiv (L,y), \quad
  {}^\forall \!y \in [0,L] , \label{z4}
\end{equation}
which is closely related to the orbifold compactification on $T^2/Z_4$.
We include a gauge field as well as a chiral fermion in the bulk.
Here we argue the bulk fermion with positive chirality.
It is straightforward to extend the arguments to fermions with
negative chirality.

Let us study the Lagrangian,
\begin{equation}
  {\cal L} = {\cal L}_{\psi_+^{}} + {\cal L}_{\rm gauge}
\end{equation}
with
\begin{equation}
  {\cal L}_{\psi_+^{}} = \bar{\psi}_+ iD_M \Gamma^M \psi_+
\end{equation}
and
\begin{equation}
  {\cal L}_{\rm gauge} =  - \frac{1}{4}F_{MN}^a F^{a\,MN}
\end{equation}
where $M=0,1,2,3,5,6$,
\begin{equation}
  D_M \equiv \frac{1}{2}\DLR_M - i g_{6D}^{} A_M,
\end{equation}
and
\begin{equation}
  F_{MN}^a \equiv \partial_M A_N^a - \partial_N A_M^a
  + g_{6D}^{} f^{abc} A_M^b A_N^c .
\end{equation}
The chiral fermions $\psi_\pm$ in the bulk are defined as
\begin{equation}
  \psi_\pm \equiv P_\pm \psi,
\end{equation}
with the chiral projection operators $P_\pm$,
\begin{equation}
  P_\pm \equiv \frac{1}{2}\left(1 \pm \Gamma_{\chi,7} \right),
\end{equation}
where the chirality matrix $\Gamma_{\chi,7}$ in six dimensions is
\begin{equation}
  \Gamma_{\chi,7}\equiv
  \Gamma^0 \Gamma^1 \Gamma^2 \Gamma^3 \Gamma^5 \Gamma^6, \quad
  \Gamma_{\chi,7} \Gamma_{\chi,7} = 1.
\end{equation}
Hence the fermions $\psi_\pm$ have four components.
For our purpose, it is convenient to use four dimensional
right/left-handed notations.
The four dimensional chirality matrix $\Gamma_{\chi,5}$ is
\begin{equation}
 \Gamma_{\chi,5}\equiv i \Gamma^0 \Gamma^1 \Gamma^2 \Gamma^3 ,\quad
 \Gamma_{\chi,5} \Gamma_{\chi,5}=1.
\end{equation}
The matrices $\Gamma_{\chi,5}$ and $\Gamma_{\chi,7}$ satisfy
\begin{equation}
 [\Gamma_{\chi,5},\Gamma_{\chi,7}] = 0,
\end{equation}
so that $\Gamma_{\chi,5}$ and $\Gamma_{\chi,7}$ are simultaneously
diagonalizable.
Thus we further decompose $\psi_\pm$ into the four dimensional
right/left-handed fermions:
\begin{equation}
  \psi_\pm = \psi_{\pm R} + \psi_{\pm L},
\end{equation}
where
\begin{equation}
  \psi_{\pm R} \equiv P_R \psi_{\pm}, \quad
  \psi_{\pm L} \equiv P_L \psi_{\pm},
\end{equation}
with the four dimensional chiral projection operators $P_{R,L}$,
\begin{equation}
  P_{R,L} \equiv \frac{1}{2}\left(1 \pm \Gamma_{\chi,5} \right).
\end{equation}
Noting
\begin{equation}
  \{\Gamma^\mu,\Gamma_{\chi,5}\}=0, \quad \mbox{for} \quad
  \mu = 0,1,2,3
\end{equation}
and
\begin{equation}
  [\Gamma^m,\Gamma_{\chi,5}]=0, \quad \mbox{for} \quad
  m=5,6,
\end{equation}
the Lagrangian ${\cal L}_{\psi_+^{}}$ is rewritten in terms of
$\psi_{+R}$ and $\psi_{+L}$ as follows:
\begin{equation}
  {\cal L}_{\psi_+^{}} = {\cal L}_{RR+LL}^{} + {\cal L}_{RL+LR}^{},
\end{equation}
with
\begin{equation}
{\cal L}_{RR+LL}^{}  \equiv
  \bar{\psi}_{+R} iD_\mu \Gamma^\mu \psi_{+R}
+ \bar{\psi}_{+L} iD_\mu \Gamma^\mu \psi_{+L},
  \label{Lag-psi-RR}
\end{equation}
and
\begin{equation}
{\cal L}_{RL+LR}^{} \equiv
  \bar{\psi}_{+R} iD_m \Gamma^m \psi_{+L}
+ \bar{\psi}_{+L} iD_m \Gamma^m \psi_{+R}.
  \label{Lag-psi-RL}
\end{equation}

Under the identification (\ref{z4}), the Lagrangian should be the same:
\begin{equation}
  {\cal L}|^{(y,0)} = {\cal L}|^{(0,y)}, \quad
  {\cal L}|^{(y,L)} = {\cal L}|^{(L,y)}. \label{Lag_z4}
\end{equation}
We then impose the BC's on fermions as
\begin{subequations}
\label{f-RL-z4}
\begin{align}
  \psi_{+R}(y,0) &= e^{\frac{i\pi}{2}n}\psi_{+R}(0,y),
  \label{f-R-z4}
  \\
  \psi_{+L}(y,0) &= ie^{\frac{i\pi}{2}n}\psi_{+L}(0,y),
  \label{f-L-z4}
\end{align}
\end{subequations}
and
\begin{subequations}
\label{f-RL2-z4}
\begin{align}
  \psi_{+R}(y,L) &= (-1)^\ell e^{\frac{i\pi}{2}n}\psi_{+R}(L,y),
  \label{f-R2-z4}
  \\
  \psi_{+L}(y,L) &= i\,(-1)^\ell e^{\frac{i\pi}{2}n}\psi_{+L}(L,y),
  \label{f-L2-z4}
\end{align}
\end{subequations}
where the integers $n$ and $\ell$ can take the values of
$n=0,1,2,3$ and $\ell =0,1$, respectively.
Differentiating the BC's (\ref{f-RL-z4})--(\ref{f-RL2-z4})
with respect to $y$, we find
\begin{subequations}
\label{df-RL-z4}
\begin{align}
  \partial_5 \psi_{+R}(y,0) &=
  e^{\frac{i\pi}{2}n} \partial_6 \psi_{+R}(0,y),
  \label{df-R-z4}
  \\
  \partial_5 \psi_{+L}(y,0) &=
  ie^{\frac{i\pi}{2}n} \partial_6 \psi_{+L}(0,y),
  \label{df-L-z4}
\end{align}
\end{subequations}
and
\begin{subequations}
\label{df-RL2-z4}
\begin{align}
  \partial_5 \psi_{+R}(y,L) &=
  (-1)^\ell e^{\frac{i\pi}{2}n} \partial_6 \psi_{+R}(L,y),
  \label{df-R2-z4}
  \\
  \partial_5 \psi_{+L}(y,L) &=
  i\,(-1)^\ell e^{\frac{i\pi}{2}n} \partial_6 \psi_{+L}(L,y).
  \label{df-L2-z4}
\end{align}
\end{subequations}
We further impose the BC's on the derivative terms as
\begin{subequations}
\label{df2-RL-z4}
\begin{align}
  \partial_6 \psi_{+R}(y,0) &=
  -e^{\frac{i\pi}{2}n} \partial_5 \psi_{+R}(0,y),
  \label{df2-R-z4}
  \\
  \partial_6 \psi_{+L}(y,0) &=
  -ie^{\frac{i\pi}{2}n} \partial_5 \psi_{+L}(0,y),
  \label{df2-L-z4}
\end{align}
\end{subequations}
and
\begin{subequations}
\label{df2-RL2-z4}
\begin{align}
  \partial_6 \psi_{+R}(y,L) &=
  (-1)^{\ell+1} e^{\frac{i\pi}{2}n} \partial_5 \psi_{+R}(L,y),
  \label{df2-R2-z4}
  \\
  \partial_6 \psi_{+L}(y,L) &=
  i\,(-1)^{\ell+1} e^{\frac{i\pi}{2}n} \partial_5 \psi_{+L}(L,y).
  \label{df2-L2-z4}
\end{align}
\end{subequations}
The BC's of the derivative terms imply the identification of
gauge bosons as
\begin{subequations}
\label{gauge-z4}
\begin{align}
  A_\mu (y,0) &= A_\mu (0,y), & A_\mu (y,L) &= A_\mu (L,y),
  \label{g-z4}
  \\
  A_{5} (y,0) &= A_{6}(0,y),  & A_{5} (y,L) &= A_{6}(L,y),
  \label{g5-z4}
  \\
  A_{6} (y,0) &= -A_{5}(0,y), & A_{6} (y,L) &= -A_{5}(L,y).
  \label{g6-z4}
\end{align}
\end{subequations}
We differentiate Eq.~(\ref{gauge-z4}) with respect to $y$ and find
\begin{subequations}
\label{dgauge-z4}
\begin{eqnarray}
  \partial_5 A_\mu |^{(y,0),\,(y,L)} &=& \partial_6 A_\mu |^{(0,y),\,(L,y)},
  \label{dg-z4}
  \\[2mm]
  \partial_5 A_{6}|^{(y,0),\,(y,L)} &=& - \partial_6 A_{5}|^{(0,y),\,(L,y)}.
  \label{dg6-z4}
\end{eqnarray}
\end{subequations}
The identification (\ref{Lag_z4}) for the gauge sector ${\cal L}_{\rm gauge}$
then requires the BC's
\begin{subequations}
\label{dgauge2-z4}
\begin{eqnarray}
  \partial_6 A_\mu |^{(y,0),\,(y,L)} &=& -\partial_5 A_\mu |^{(0,y),\,(L,y)},
  \label{dg2-z4}
  \\[2mm]
  \partial_6 A_{5}|^{(y,0),\,(y,L)} &=& - \partial_5 A_{6}|^{(0,y),\,(L,y)}.
  \label{dg5-z4}
\end{eqnarray}
\end{subequations}

Now it is easy to check that the identification (\ref{Lag_z4}) is
satisfied. From the BC's (\ref{f-RL-z4})--(\ref{f-RL2-z4}),
${\cal L}_{RR+LL}^{}$ defined by Eq.~(\ref{Lag-psi-RR})
is obviously identical to the reflection under Eq.~(\ref{z4}).
To see the identity for ${\cal L}_{RL+LR}^{}$,
we apply the relations
\begin{equation}
  \Gamma^5 P_R P_\pm = \pm i\Gamma^6 P_R P_\pm, \quad
  \Gamma^5 P_L P_\pm = \mp i\Gamma^6 P_L P_\pm,
  \label{gamma5-6}
\end{equation}
and then find
\begin{eqnarray}
{\cal L}_{RL+LR}^{} &=& \phantom{+}
  \bar{\psi}_{+R} (D_5\Gamma^6 - D_6 \Gamma^5) \psi_{+L}
  \nonumber \\ && +
  \bar{\psi}_{+L} (-D_5\Gamma^6 + D_6 \Gamma^5) \psi_{+R}.
  \label{Lag-psi-RL2}
\end{eqnarray}
By using the BC's of Eqs.~(\ref{f-RL-z4})--(\ref{df2-RL2-z4}),
we can easily confirm the identification
${\cal L}_{\psi_+^{}}\,|^{(y,0),\,(y,L)} =
 {\cal L}_{\psi_+^{}}\,|^{(0,y),\,(L,y)}$.
How about the identification of the gauge sector?
The derivative of Eq.~(\ref{gauge-z4}) with respect to $x^\mu$ and
Eqs.~(\ref{dgauge-z4})--(\ref{dgauge2-z4}) yield
\begin{subequations}
\label{F-z4}
\begin{eqnarray}
 F_{\mu\nu}^a\,|^{(y,0),\,(y,L)} &=& \phantom{-}
 F_{\mu\nu}^a\,|^{(0,y),\,(L,y)},\\
 F_{\mu 5}^a\,|^{(y,0),\,(y,L)} &=& \phantom{-}
 F_{\mu 6}^a\,|^{(0,y),\,(L,y)},\\
 F_{\mu 6}^a\,|^{(y,0),\,(y,L)} &=& -F_{\mu 5}^a\,|^{(0,y),\,(L,y)},\\
 F_{56}^a\,|^{(y,0),\,(y,L)} &=& \phantom{-}
 F_{56}^a\,|^{(0,y),\,(L,y)},
\end{eqnarray}
\end{subequations}
so that the identification ${\cal L}_{\rm gauge}\,|^{(y,0),\,(y,L)} =
{\cal L}_{\rm gauge}\,|^{(0,y),\,(L,y)}$ is clearly satisfied.

We can show that the phase factor should be
$e^{\frac{i\pi}{2}n}$ $(n=0,1,2,3)$
after some algebraic calculation.~\cite{Dobrescu:2004zi}
We will not repeat it here.
In this paper, we take the convention
\begin{equation}
  n=0, \quad \ell = 0 .
\end{equation}

The BC's yield the Kaluza-Klein (KK) decomposition of
the gauge field $A_\mu$:
\begin{eqnarray}
  A_\mu(x^\mu,y^5,y^6) &=&
  \frac{1}{L} \sum_{j \geq k \geq 0} A_{\mu,cc}^{[j, k]}(x^\mu)
  f_{cc}^{[j,k]} \nonumber \\ &+&
  \frac{1}{L} \sum_{j > k > 0} A_{\mu,ss}^{[j, k]}(x^\mu)
  f_{ss}^{[j,k]} ,
\end{eqnarray}
with
\begin{equation}
f_{cc}^{[0,0]} \equiv 1, \quad
f_{cc}^{[j \ne 0,0]} \equiv
  \cos\left(\dfrac{\pi}{L} j y^5\right)
+ \cos\left(\dfrac{\pi}{L} j y^6\right),
\end{equation}
\begin{equation}
f_{cc}^{[j,j]} \equiv
  2\cos\left(\dfrac{\pi}{L} j y^5\right)
   \cos\left(\dfrac{\pi}{L} j y^6\right), \; (j \ne 0)
\end{equation}
\begin{eqnarray}
f_{cc}^{[j,k]} &\equiv& \sqrt{2} \left[\,
  \cos\left(\dfrac{\pi}{L} j y^5\right)
  \cos\left(\dfrac{\pi}{L} k y^6\right)
  \right. \nonumber \\ && \quad \left.
 +\cos\left(\dfrac{\pi}{L} k y^5\right)
  \cos\left(\dfrac{\pi}{L} j y^6\right) \,\right], \;(j > k > 0)
  \nonumber \\[2mm] &&
\end{eqnarray}
and
\begin{eqnarray}
  f_{ss}^{[j,k]} &\equiv& - \sqrt{2}\left[\,
  \sin\left(\dfrac{\pi}{L} j y^5\right)
  \sin\left(\dfrac{\pi}{L} k y^6\right)
  \right. \nonumber \\ && \quad \left.
 -\sin\left(\dfrac{\pi}{L} k y^5\right)
  \sin\left(\dfrac{\pi}{L} j y^6\right)\,\right].
\end{eqnarray}

The KK expansions of $\psi_{+R}$ and $\psi_{+L}$ are given by
\begin{align}
  \psi_{+R}(x^\mu,y^5,y^6) &=
  \frac{1}{L} \sum_{j \geq k \geq 0} \psi_{+R,cc}^{[j, k]}(x^\mu)
  f_{cc}^{[j,k]} \nonumber \\ &+
  \frac{1}{L} \sum_{j > k > 0} \psi_{+R,ss}^{[j, k]}(x^\mu)
  f_{ss}^{[j,k]} ,
\end{align}
and
\begin{align}
  \psi_{+L}(x^\mu,y^5,y^6) &=
  \frac{1}{L} \sum_{j  \geq k \geq 0} \psi_{+L,1}^{[j, k]}(x^\mu)
  g_1^{[j,k]} \nonumber \\ &+
  \frac{1}{L} \sum_{j > k > 0} \psi_{+L,2}^{[j, k]}(x^\mu)
  g_2^{[j,k]} ,
\end{align}
with
\begin{equation}
  g_1^{[j,k]} \equiv \frac{j}{\sqrt{j^2+k^2}}f_{sc}^{[j,k]}
  -\frac{ik}{\sqrt{j^2+k^2}}f_{cs}^{[j,k]},
\end{equation}
and
\begin{equation}
  g_2^{[j,k]} \equiv -\frac{ik}{\sqrt{j^2+k^2}}f_{sc}^{[j,k]}
  +\frac{j}{\sqrt{j^2+k^2}}f_{cs}^{[j,k]} ,
\end{equation}
where
\begin{eqnarray}
f_{sc}^{[j,k]} &\equiv& \sqrt{\frac{2}{1+\delta_{k,0}}} \left[\,
  \sin\left(\dfrac{\pi}{L} j y^5\right)
  \cos\left(\dfrac{\pi}{L} k y^6\right)
  \right. \nonumber \\ && \quad \left.
 -i\cos\left(\dfrac{\pi}{L} k y^5\right)
  \sin\left(\dfrac{\pi}{L} j y^6\right) \,\right],
\end{eqnarray}
and
\begin{eqnarray}
f_{cs}^{[j,k]} &\equiv& \sqrt{\frac{2}{1+\delta_{j,0}}} \left[\,
  \cos\left(\dfrac{\pi}{L} j y^5\right)
  \sin\left(\dfrac{\pi}{L} k y^6\right)
  \right. \nonumber \\ && \quad \left.
 +i\sin\left(\dfrac{\pi}{L} k y^5\right)
  \cos\left(\dfrac{\pi}{L} j y^6\right) \,\right] .
\end{eqnarray}
The KK decompositions of $\psi_{-L}$ and $\psi_{-R}$ are
the same as those of $\psi_{+R}$ and $\psi_{+L}$:
\begin{align}
  \psi_{-L}(x^\mu,y^5,y^6) &=
  \frac{1}{L} \sum_{j \geq k \geq 0} \psi_{-L,cc}^{[j, k]}(x^\mu)
  f_{cc}^{[j,k]} \nonumber \\ &+
  \frac{1}{L} \sum_{j > k > 0} \psi_{-L,ss}^{[j, k]}(x^\mu)
  f_{ss}^{[j,k]} ,
\end{align}
and
\begin{align}
  \psi_{-R}(x^\mu,y^5,y^6) &=
  \frac{1}{L} \sum_{j  \geq k \geq 0} \psi_{-R,1}^{[j, k]}(x^\mu)
  g_1^{[j,k]} \nonumber \\ &+
  \frac{1}{L} \sum_{j > k > 0} \psi_{-R,2}^{[j, k]}(x^\mu)
  g_2^{[j,k]} .
\end{align}

In this way, the chiral compactification on the square leaves
the zero modes $A_{\mu,cc}^{[0,0]},\psi_{+R,cc}^{[0,0]}$, and
$\psi_{-L,cc}^{[0,0]}$.

\section{Gauge symmetry on boundaries}
\label{app-b}

We study the gauge symmetry on the boundaries.
The four/five dimensional notations are more convenient than the six
dimensional one.
We thus rewrite ${\cal L}_{\psi_+^{}}$ by using the following
representation of the gamma matrices,
\begin{eqnarray}
 \Gamma^\mu &=&
 \left(\begin{array}{cc}\gamma^\mu & 0 \\ 0 & \gamma^\mu
 \end{array}\right), \\[3mm]
 \Gamma^5 &=&
 \left(\begin{array}{cc}0 & i\gamma_5 \\ i\gamma_5 & 0
 \end{array}\right), \\[3mm]
 \Gamma^6 &=&
 \left(\begin{array}{cc}0 & \gamma_5 \\ -\gamma_5 & 0
 \end{array}\right),
\end{eqnarray}
where $\gamma^\mu$ and $\gamma_5$ are 4$\times$4 matrices.
We take the space-time metric to ${\rm diag}(+,-,-,\cdots,-)$,
so that the five dimensional gamma matrices are $\gamma^\mu,i\gamma_5$.
Noting that
\begin{equation}
 \Gamma_{\chi, 7} \equiv
 \Gamma^0 \Gamma^1 \Gamma^2 \Gamma^3 \Gamma^5 \Gamma^6 =
 \left(\begin{array}{cc} -\gamma_5 & 0 \\ 0 & \gamma_5
 \end{array}\right),
\end{equation}
the chiral fermions $\psi_{+L,+R}$ should be
\begin{equation}
  \psi_{+L} \to
  \left(\begin{array}{@{}c@{}} \psi_{+L} \\ 0 \end{array}\right), \quad
  \psi_{+R} \to
  \left(\begin{array}{@{}c@{}} 0 \\ \psi_{+R} \end{array}\right).
\end{equation}
In the four/five dimensional notations, the Lagrangian
${\cal L}_{\psi_+^{}} (= {\cal L}_{RR+LL}^{} + {\cal L}_{RL+LR}^{})$
is represented as
\begin{equation}
{\cal L}_{RR+LL}^{}  =
  \bar{\psi}_{+R} iD_\mu \gamma^\mu \psi_{+R}
+ \bar{\psi}_{+L} iD_\mu \gamma^\mu \psi_{+L},
  \label{Lag-psi-RR4}
\end{equation}
and
\begin{eqnarray}
{\cal L}_{RL+LR}^{} & = & \phantom{+}
  \bar{\psi}_{+R} [iD_5 - D_6 ] (i\gamma_5) \psi_{+L}
 \nonumber \\ &&
+ \bar{\psi}_{+L} [iD_5 + D_6 ] (i\gamma_5) \psi_{+R}.
  \label{Lag-psi-RL4}
\end{eqnarray}
By performing integration by parts, we obtain the $RL+LR$ part of the action,
\begin{eqnarray}
\lefteqn{ \hspace*{-4mm}
 S_{RL+LR}^{} = \int dx^4 \int dy^5 dy^6 {\cal L}'_{RL+LR}
} \nonumber \\[2mm] && \hspace*{-3mm}
+ \frac{1}{2}\int dx^4 \int dy^6 \Bigg[\,
 (\bar{\psi}_{+R}\psi_{+L}) + \mbox{(h.c.)}
 \,\Bigg]_{(0,y^6)}^{(L,y^6)}  \nonumber \\[2mm] && \hspace*{-3mm}
+ \frac{1}{2}\int dx^4 \int dy^5 \Bigg[\,
 i(\bar{\psi}_{+R}\psi_{+L}) + \mbox{(h.c.)}
 \,\Bigg]_{(y^5,0)}^{(y^5,L)}
\label{S-RL}
\end{eqnarray}
with
\begin{eqnarray}
\lefteqn{ \hspace*{-5mm}
{\cal L}'_{RL+LR} \equiv
(-i\partial_5 + \partial_6)\bar{\psi}_{+R} (i\gamma_5) \psi_{+L}
} \nonumber \\ && \hspace*{-3mm}
+\bar{\psi}_{+R} [g_{6D}^{}A_5+ig_{6D}^{}A_6](i\gamma_5) \psi_{+L}
\nonumber \\ && \hspace*{-3mm} +
\bar{\psi}_{+L} (i\partial_5 + \partial_6 + g_{6D}^{}A_5-ig_{6D}^{}A_6)
 (i\gamma_5) \psi_{+R} .
\label{L-RL}
\end{eqnarray}
The surface terms in Eq.~(\ref{S-RL}) are vanishing
thanks to the BC's (\ref{f-RL-z4})--(\ref{f-RL2-z4}).
Therefore we may use ${\cal L}'_{RL+LR}$ instead of ${\cal L}_{RL+LR}$.

Now we impose the BC's in order to restore the five dimensional gauge
symmetry on the boundaries.
Since the gauge scalars should be massive,
it is natural to assign the Dirichlet BC's to $A_5$ and $A_6$, i.e.,
\begin{equation}
  A_5\,|^{(0,y),(L,y)} = 0, \quad A_6\,|^{(y,0),(y,L)} = 0.
\end{equation}
Then the derivative term of $\psi_{+R}$ should be zero simultaneously from
Eq.~(\ref{L-RL}),
\begin{equation}
  \partial_5 \psi_{+R}\,|^{(0,y),(L,y)} = 0 , \quad
  \partial_6 \psi_{+R}\,|^{(y,0),(y,L)} = 0 .
\end{equation}
In order to ensure nonvanishing ${\cal L}'_{RL+LR}$,
we impose
\begin{equation}
  (\partial_5 + i \partial_6) \psi_{+L}\,|^{(0,y),(L,y),(y,0),(y,L)} \ne 0.
\end{equation}
The BC's for the gauge bosons are easily found as
\begin{equation}
  \partial_5 A_\mu\,|^{(0,y),(L,y)} = 0 , \quad
  \partial_6 A_\mu\,|^{(y,0),(y,L)} = 0 ,
\end{equation}
and
\begin{equation}
  \partial_5 A_6\,|^{(0,y),(L,y)} = 0 , \quad
  \partial_6 A_5\,|^{(y,0),(y,L)} = 0 .
\end{equation}

We did not fix the gauge yet, so that the mixing terms of 
$A_\mu A_5$ and $A_\mu A_6$ remain.
For completeness, one may introduce $R_\xi$ gauge fixing terms, etc..

\end{document}